\documentstyle[prl,aps,amsfonts]{revtex}
\begin{document}

\title{$q$-deformed dynamics of $q$-deformed oscillators}
\author{Jos\'e Luis Gruver\thanks{E-mail: gruverj@mail.biu.ac.il}}
\address{Department of Physics and Jack and Pearl Resnick Institute of 
Advanced Technology, Bar-Ilan University, Ramat-Gan 52900, Israel}

\maketitle

\begin{abstract}
We show that an infinite set of $q$-deformed relevant operators
close a partial $q$-deformed Lie algebra under commutation with the 
Arik-Coon oscillator. The dynamics is described by the multicommutator: 
$\lbrack \hat H, \ldots, \lbrack \hat H, \hat O \rbrack \ldots \rbrack$, 
that follows a power law which leads to a dynamical scaling. 
We study the dynamics of the Arik-Coon and anharmonic oscillators
and analyze the role of $q$ and the other parameters in the evolution
of both systems.
\\[0.3truecm]
\\[0.3truecm]
\end{abstract}

Quantum algebras, also known as quantum groups, have been the subject
of an intensively research in the last years 
\cite{Monteiro,Chung,Pan,Macfarlane,Biedenharn,Bonatsos1,Bonatsos2,Helio4,Chaichian,Tombesi,Bonatsos3,Mancini,Celeghini,Tsallis,Abe,Arik}.
Particularly, after the works of Macfarlane \cite{Macfarlane} and 
Biedenharn \cite{Biedenharn} on the $q$-deformed oscillators, a great 
effort has been devoted to the application and generalization of 
$q$-deformed systems in chemistry and physics. They have been used as
a model to describe vibration of polyatomic molecules \cite{Bonatsos1}, 
to study vortices in superfluid films \cite{Bonatsos2}, and 
to analyze the phonon spectrum in $^{4}$He \cite{Helio4}.
In quantum optics,
$q$-bosons have been used to generalized fundamental models such us, 
the Jaynes-Cummings \cite{Chaichian} and Dicke models \cite{Tombesi}. 
Besides, using generalized deformed oscillators several versions of the
Jaynes-Cummings Hamiltonian have found a unified description
\cite{Bonatsos3}. Finally, we can mention that, the important concepts 
of $f$-coherent \cite{Mancini} and squeezed states \cite{Celeghini} 
have been also treated in the frame of $q$-deformed theories. 

Recently, the possible relation between nonextensive statistical mechanics 
\cite{Tsallis} and quantum groups has renewed the interest on $q$-deformed 
systems, particularly, due to the nonextensivity properties inherent to
$q$-deformed theories \cite{Tsallis,Abe,Arik}. Moreover, since
$q$-deformed systems, under certain conditions, resemble the features 
of the nonlinear ones, they are potential candidates for studying 
nonlinear problems where the nonextensivity is a desired property. 

The purpose of the present work is to study the dynamical properties of 
$q$-deformed oscillators and their relationship to the anharmonic 
oscillators by means of a Lie-algebraic approach. In doing so, we find 
that an infinite dimensional set of ``$q$-deformed relevant operators''
close a ``partial $q$-deformed Lie algebra'' under commutation with 
the Arik-Coon Hamiltonian. We show that the dynamics of the
system can be described in terms of the multicommutator of the type 
$\lbrack \hat H, \ldots, \lbrack \hat H, \hat O \rbrack \ldots \rbrack$. 
We also obtain, that the multicommutator can be expressed for $q > 1$
as an {\it operator average} with respect to the ``Binomial distribution'' 
which depends {\it only}
on the deformation parameter $q$, and for the general case (i.e. $q
\in \Re$) as a ``power law''. As a consequence of the power law
dependence, 
we find that the dynamics of the infinite-dimensional $q$-deformed 
Lie-algebra scale, i.e. the temporal evolution for the whole set of 
relevant operators collapse on a single curve. We calculate and 
analyze, the temporal evolution of the set of relevant operators 
for the $q$-deformed and the anharmonic oscillator when the initial 
conditions are a $q$-coherent and coherent states respectively.
We obtain that the dynamics of both models is governed by a 
{\it weighted average} with respect to the ``$q$-deformed Poisson'' 
and the ``standard Poisson'' distributions respectively.
Finally, we find the conditions under which the dynamics of the
relevant operators of both oscillators are isomorphous, and we
conclude that $q$ is not only related to the anharmonicity parameter
but also depends on the index that characterizes the relevant operators.

Before going into the core of our Letter, let us briefly review the 
basis of our approach. As shown by Alhassid and Levine \cite{alh}, the
essential point for the description of a system's temporal evolution is to
be found in the closing of a partial Lie algebra with the Hamiltonian 
under scrutiny, in the fashion 

\begin{equation}
\label{cierre}
\lbrack \hat H,\hat O_{k} \rbrack = i\hbar \,\sum_{j=0}^{L}\,g_{jk} \,
\hat O_{j},
\end{equation}
an equation that defines the so-called {\it set of relevant operators} 
\cite{RO}, or {\it observation level} \cite{alem}. Notice that $L$
can be finite or infinite, depending on the nature of the Hamiltonian 
at hand \cite{nos}. 
Given an observation level whose operators do not explicitly
depend upon the time, the relevant dynamics can be obtained in terms 
of multicommutarors of the type:

\begin{equation}
\label{serie11}
\underbrace{\lbrack \hat H,\dots, \lbrack \hat H, \hat O_{k}
\rbrack \dots \rbrack}_{j \; \mbox{times}},
\end{equation}
as follow:
\begin{equation}
\label{serie} 
\hat O_{k}(t) = \sum_{j=0}^{\infty} \frac{(it/\hbar)^{j}}{j!}
\underbrace{\lbrack \hat H,\dots, \lbrack \hat H, \hat O_{k}
\rbrack \dots \rbrack}_{j \; \mbox{times}}.
\end{equation}

Equation (\ref{serie}), besides of describing the temporal evolution, 
tells us how many and which are the relevant operators involved up a given 
order {\it j} in time. Therefore, the multicommutator fully describe the 
development of the correlations at different orders in time \cite{Gruver}.

 From now onwards, we will be concerned with the Arik-Coon Hamiltonian 
\cite{Arik-Coon}

\begin{equation}
\label{q-osc}
\hat H_{q} = \hbar \omega_{q} \hat \Delta_{q},
\end{equation}
$\omega_{q}$ stands for the oscillator energy
and $\hat \Delta_{q} \equiv \hat a_{q}^{\dagger} \hat a_{q}$.
The creation operator $\hat a_{q}^{\dagger}$ and its hermitian conjugate 
$\hat a_{q}$ satisfy: $\hat a_{q} \hat a_{q}^{\dagger} = 
q \hat a_{q}^{\dagger} \hat a_{q} + 1$ with $q \in \Re$. Besides, 
$\hat a_{q}^{\dagger} |n\rangle = \sqrt{[n+1]_{q}} |n+1\rangle$ and 
$\hat a_{q} |n\rangle = \sqrt{[n]_{q}} |n-1\rangle$, then
$\hat H_{q} |n\rangle = E_{q}(n) |n\rangle$ and 
$E_{q}(n) \equiv \hbar \omega_{q} [n]_{q}$ is the energy spectrum of 
$\hat H_{q}$. Finally, $[n]_{q} \equiv (q^{n}-1)/(q-1)$,
$[n]_{q}!\equiv[n]_{q}[n-1]_{q}\cdots[1]_{q}$, and $[0]_{q}!\equiv1$.

With the above definitions the commutation relation reads:

\begin{equation}
\label{fundamental}
\lbrack \hat a_{q}, \hat a_{q}^{\dagger} \rbrack = 1 + {(q-1) 
\over \hbar\omega_{q}} \hat H_{q},
\end{equation}
which turns out to be simplest generalization of the commutation relation
introduced in high energy physics \cite{Saavedra} and applied to
several physical problems \cite{Jannussis1993,Codriansky}.
  
Introducing $\{ \hat \Lambda_{q}^{n, m} \equiv (\hat a_{q}^{\dagger})^{n} 
\hat \Delta_{q}^{m} \}_{n,m=0}^{\infty}$ and their hermitian conjugates, 
it can be proved that

\begin{mathletters}
\label{relevantoperators}
\begin{eqnarray}
\hat \Lambda^{n, m}_{q,+} & \equiv &
\hat \Lambda_{q}^{n, m} + (\hat \Lambda_{q}^{n, m})^{\dagger}, \\
\hat \Lambda^{n, m}_{q,-} & \equiv & i
\lbrack \hat \Lambda_{q}^{n, m} - (\hat \Lambda_{q}^{n, m})^{\dagger} \rbrack,
\end{eqnarray}
\end{mathletters}
or equivalently
\begin{mathletters}
\label{transformations}
\begin{eqnarray}
\hat \Lambda_{q}^{n, m} & = & {\hat \Lambda^{n, m}_{q,+} 
- i \hat \Lambda^{n, m}_{q,-} \over 2}, \\ 
(\hat \Lambda_{q}^{n, m})^{\dagger} & = &
{\hat \Lambda^{n, m}_{q,+} + i \hat \Lambda^{n, m}_{q,-} \over 2},
\end{eqnarray}
\end{mathletters}
close a ``partial $q$-deformed Lie algebra'' under commutation with the 
Hamiltonian (\ref{q-osc}) of the form:

\begin{mathletters}
\label{partialalgebra}
\begin{eqnarray}
\lbrack \hat H_{q}, \hat \Lambda_{q}^{n, m} \rbrack & = & 
E_{q}(n) \hat \Lambda_{q}^{n, m} + 
E_{q}(n) (q-1) \hat \Lambda_{q}^{n, m+1}, \\
\lbrack \hat H_{q}, (\hat \Lambda_{q}^{n, m})^{\dagger} \rbrack & = &
- E_{q}(n) (\hat \Lambda_{q}^{n, m})^{\dagger} -
E_{q}(n) (q-1) (\hat \Lambda_{q}^{n, m+1})^{\dagger}.
\end{eqnarray}
\end{mathletters}

Let us note that our partial Lie algebras are decoupled with respect to the 
supra-index $n$. For $n=1$ we have the 
``$q$-deformed Heisenberg-like-infinite group'' (i.e., $\{\hat
a_{q}^{\dagger} 
\hat \Delta_{q}^{m} \}_{m=0}^{\infty}$). 
It is important to mention that any $q$-deformed relevant operator can
be written in terms of the basis defined by Eqs. (\ref{relevantoperators}) 
or (\ref{transformations}), for instance we can use them to study the 
evolution of the squeezed or the $n$th-coherence operators \cite{Meystre}.

It is interesting to mention that, for $|q| < 1$ the relevant operators
$\hat \Lambda_{q}^{n, m}$ are bounded and for 
$|q| > 1$ they are not, while the set of relevant operators associated 
with the anharmonic oscillator are unbounded. As we will see
later, this property is consistent with the fact, that {\it only} for 
$q > 1$ there is a regime of parameters such that the set of relevant 
operators of both models are dynamically isomorphous.
 
As it is expressed in Eq. (\ref{serie}) the multicommutator constitutes the
main tool in order to evaluate the temporal evolution of our set
of relevant operators. Before studying the most general case 
(i.e. $q \in \Re$) let us evaluate the multicommutator for $q > 1$ 
then, after some algebra, we arrive to a compact expression for the
multicommutator, namely

\begin{equation}
\label{multicommutator0}
\underbrace{\lbrack \hat H,\dots, \lbrack 
\hat H_{q}, \hat \Lambda_{q}^{n,m}  
\rbrack \dots \rbrack}_{j \; \mbox{times}} =
Z_{[n]_{q}}^{j} \sum_{k=0}^{j} B(j,k,q^{-1}) \hat \Lambda_{q}^{n,m+k},
\end{equation}
$Z_{[n]_{q}} \equiv E_{q}(n)q$ and $B(j,k,q^{-1}) \equiv \biggl(
\begin{array}{c} j \\ k \end{array}
\biggr) (q^{-1})^{j-k}(1-q^{-1})^{k}$ is the ``Binomial distribution'' with
mean value $jq^{-1}$ and variance $jq^{-1}(1-q^{-1})$. 
Equation (\ref{multicommutator0}) express the multicommutator as an
{\it operator average with respect to the Binomial distribution
which depends only on} $q^{-1}$. Note that as $j$ goes to infinity,
$B(j,k,q^{-1})$ approaches a Gaussian distribution with mean value
and variance independent of $n$ and $m$.

Now, for general case (i.e. $q \in \Re$) we obtain:

\begin{equation}
\label{multicommutator1}
\underbrace{\lbrack \hat H,\dots, \lbrack
\hat H_{q}, \hat \Lambda_{q}^{n,m}
\rbrack \dots \rbrack}_{j \; \mbox{times}} =
\hat \Lambda_{q}^{n,m}
(E_{q}(n) \lbrack \hat a_{q}, \hat a_{q}^{\dagger} \rbrack)^{j}.
\end{equation}

Equation (\ref{multicommutator1}) express the multicommutator as integer 
powers of $E_{q}(n) \lbrack \hat a_{q}, \hat a_{q}^{\dagger} \rbrack$ 
(which is a constant of the motion) and as a function of
$\hat \Lambda_{q}^{n, m}$. Defining $\tau \equiv \omega_{q}t$, and 
after some more algebra, we arrive to the scaling formula for the
dynamics:

\begin{equation}
\label{scaling}
\left\{[\hat \Lambda^{n, m}_{q}(0)]^{-1}
\hat \Lambda^{n, m}_{q}(\tau)\right\}^{1/[n]_{q}}=
\exp(i \tau \lbrack \hat a_{q}, \hat a_{q}^{\dagger} \rbrack).
\end{equation}

To study the temporal evolution of the relevant operators introduced
in Eqs. (\ref{relevantoperators}) and (\ref{transformations}) we 
recall Eqs. (\ref{serie}) and (\ref{multicommutator0}),
so after some algebra we arrive to the following expression:

\begin{eqnarray}
\label{evo1}
\hat \Lambda_{q}^{n,m}(\tau) & = & \exp(i[n]_{q}\tau)
\sum_{r=0}^{\infty} \frac{[i[n]_{q}(q-1)\tau]^{r}}{r!}
\hat \Lambda_{q}^{n,m+r}(0).
\end{eqnarray}

This last equation can be recasted in a more appropriate way by
writing $\hat \Lambda_{q}^{n,m+r}$ in normal order. 
Normal ordering is achieved 
by noting that $\hat \Lambda_{q}^{n,m+r}$ can be written as a linear
combination of the operators $(\hat a_{q}^{\dagger})^{n+s} (\hat
a_{q})^{s}$ as follows \cite{Katriel1}:

\begin{equation}
\label{normalordering}
\hat \Lambda_{q}^{n,m+r} = \sum_{s=0}^{m+r} S_{q}^{s,m+r} 
(\hat a_{q}^{\dagger})^{n+s} (\hat a_{q})^{s},
\end{equation}
where $S_{q}^{s,m+r} \equiv \sum_{k=0}^{s} (-1)^{s-k} q^{{(s-k)^{2}-(s-k)
\over 2}} {[k]_{q}^{m+r} \over [k]_{q}! [s-k]_{q}!}$ are the ``$q$-deformed
Stirling numbers of second kind'' \cite{Katriel2}. Working out Eq. (\ref{evo1}) and 
Eq. (\ref{normalordering}) we arrive to:

\begin{eqnarray}
\label{evo2}
\hat \Lambda_{q}^{n,m}(\tau) & = & \exp(i[n]_{q}\tau)
\sum_{k=0}^{\infty} \sum_{r=0}^{\infty} (-1)^{r} q^{{r(r-1)\over 2}}
{[k]_{q}^{m} \over [k]_{q}! [r]_{q}!}\exp[i[n]_{q}(q-1)[k]_{q}\tau]
\nonumber \\ & \times & 
[\hat a_{q}^{\dagger}(0)]^{n+r+k} [\hat a_{q}(0)]^{r+k}. 
\end{eqnarray}  

Now, we can straightforwardly calculate the temporal evolution of the
relevant operators when the initial condition is, for instance, a
``$q$-deformed coherent state'' 
(i.e. $\langle \alpha_{q}(0)| (\hat a_{q}^{\dagger})^{n+r+k} (\hat 
a_{q})^{r+k}| \alpha_{q}(0) \rangle =
[\alpha_{q}^{*}(0)]^{n}|\alpha_{q}(0)|^{2(r+k)}$). Finally,

\begin{eqnarray}
\label{q-deformed-dynamics}
\langle \hat \Lambda_{q}^{n,m} \rangle_{\tau} & = & [\alpha_{q}^{*}(0)]^{n}
\exp{(i[n]_{q}\tau)} \sum_{k=0}^{\infty}
[k]_{q}^{m} P_{q} (\alpha_{q}(0),k) 
\nonumber \\ & \times &
\exp[i[n]_{q}(q-1)[k]_{q}\tau],
\end{eqnarray}
where $P_{q}(\alpha_{q}(0),k) \equiv 
{|\alpha_{q}(0)|^{2k} [\exp_{q}(|\alpha_{q}(0)|^{2})]^{-1} \over
[k]_{q}!}$
is the ``$q$-deformed Poisson distribution'' and $\exp_{q}(x) \equiv
\sum_{k=0}^{\infty}{x^{k} \over [k]_{q}!}$ is the ``$q$-deformed
exponential function''. It is worthwhile to mention that due to the 
nonlinear dependence of $k$ on $q$ there is not closed-form solution 
for the Eq. (\ref{q-deformed-dynamics}). For $\tau = 0$, 
Eq. (\ref{q-deformed-dynamics}) reduces to $\langle \hat \Lambda_{q}^{n,m}
\rangle_{0} =[\alpha_{q}^{*}(0)]^{n} \langle \hat \Lambda_{q}^{0,m} 
\rangle_{0}$,  which follows from the relation: 

\begin{equation}
\label{relation}
\sum_{k=0}^{\infty} [k]_{q}^{m} {x^{k} \over
[k]_{q}!} = \sum_{r=0}^{m} S_{q}^{r,m} x^{r} \exp_{q}(x),
\end{equation}
and for $\tau > 0$, Eq. (\ref{q-deformed-dynamics}) shows the entanglement 
of the correlations along the temporal evolution of the system.

In order to compare the dynamics of the $q$-deformed and the anharmonic
oscillators, let us review some of the pertinent results for the
comparison.
We consider for the Hamiltonian of the anharmonic oscillator of second 
order the following \cite{Yurke}:

\begin{equation}
\label{anharmonic}
\hat H = \hbar \omega_{1} \hat \Delta + \hbar \omega_{2} \hat
\Delta^{2},
\end{equation}
where $\hat \Delta \equiv \hat a^{\dagger} \hat a$, and the creation and
annihilation operators satisfy the usual commutation relation 
$\lbrack \hat a, \hat a^{\dagger} \rbrack = 1$.

As it was shown in \cite{Gruver} the set $\{ \hat \Lambda^{n, m} \equiv 
(\hat a^{\dagger})^{n} \hat \Delta^{m} \}_{n,m=0}^{\infty}$ and their
hermitian conjugates close a partial Lie algebra under commutation with
the Hamiltonian (\ref{anharmonic}) of the form

\begin{mathletters}
\label{partialalgebra1}
\begin{eqnarray}
\lbrack \hat H, \hat \Lambda^{n, m} \rbrack & = & 
\hbar (n \omega_{1}+n^{2}\omega_{2}) \hat \Lambda^{n, m} + 
\hbar 2 n \omega_{2} \hat \Lambda^{n, m+1}, \\
\lbrack \hat H, (\hat \Lambda^{n, m})^{\dagger} \rbrack & = & -
\hbar (n \omega_{1}+n^{2}\omega_{2}) (\hat \Lambda^{n, m})^{\dagger}
- \hbar 2 n \omega_{2} (\hat \Lambda^{n, m+1})^{\dagger}.
\end{eqnarray}
\end{mathletters}

Notice that the functional form of Eqs. (\ref{partialalgebra}) and 
(\ref{partialalgebra1}) is the same, i.e. in both cases the result of
commuting an operator with the Hamiltonian augment in one unit the
supra-index $m$, in the case of the anharmonic oscillator this is a consequence
of the nonlinear term while for the $q$-oscillator this is a consequence
of the commutation relation (\ref{fundamental}). In other
words, for both systems we obtain the same algebraic structures being
the difference in the dependence on $n$ of their structure's coefficients.
Defining $p_{n}\equiv {{\omega_{1} \over \omega_{2}} + n \over
{\omega_{1} \over \omega_{2}} + n+2 }$ we obtain,

\begin{eqnarray}
\label{multicommutator2}
\underbrace{\lbrack \hat H,\dots, \lbrack
\hat H, \hat \Lambda^{n,m}
\rbrack \dots \rbrack}_{j \; \mbox{times}} & = &
Z_{n}^{j} \sum_{k=0}^{j} B(j,k,p_{n}) \hat \Lambda^{n,m+k},
\end{eqnarray}
$Z_{n} \equiv \hbar n[\omega_{1}+(n+2)\omega_{2}]$ and $B(j,k,p_{n})
\equiv
\left( \begin{array}{c} j \\ k \end{array} \right)
p_{n}^{j-k}(1-p_{n})^{k}$ is the ``Binomial distribution'' with mean
value $jp_{n}$ and variance $jp_{n}(1-p_{n})$. Note that
since $p_{n}$ depends on $n$, as $n$ grows the center of the
distribution
moves to higher values and the distribution spreads in contrary to the
$q$-deformed case which does not depend on $n$. 
If now we consider an initially coherent state the 
evolution of the relevant operators is:

\begin{eqnarray}
\label{sola1}
\langle \hat \Lambda^{n,m} \rangle_{t} & = & [\alpha^{*}(0)]^{n}
\exp{[i(n\omega_{1}+n^{2}\omega_{2})t]}
\sum_{k=0}^{\infty}
k^{m} P(\alpha(0),k)
\nonumber \\ & \times &
\exp{(i2n\omega_{2}kt)},
\end{eqnarray}
where $P(\alpha(0),k) \equiv
{|\alpha(0)|^{2k} [\exp(|\alpha(0)|^{2})]^{-1} \over k!}$
is the ``Poisson distribution''. 
Notice that Eqs. (\ref{q-deformed-dynamics}) and (\ref{sola1}) are
similar {\it only} in what respect to their functional forms. This is because,
although the $q$-deformed Poisson distribution approaches to the Poisson
distribution and $[k]_{q}$ to $k$ in the limit $q=1$, Eq. (\ref{sola1})
is {\it not} the $q=1$-limit of Eq. (\ref{q-deformed-dynamics}).

Using Eq. (\ref{relation}) for $q=1$, Eq. (\ref{sola1}) can be simplified to:

\begin{eqnarray}
\label{sola2}
\langle \hat \Lambda^{n,m} \rangle_{t} & = &
[\alpha^{*}(0)]^{n} 
\exp[i(n\omega_{1}+n^{2}\omega_{2})t]
\exp[|\alpha(0)|^{2}(\exp(i2n\omega_{2}t)-1)]
\nonumber \\ & \times &
\sum_{r=0}^{m} S^{r,m} |\alpha(0)|^{2r} 
\exp(i2n\omega_{2}rt)
\end{eqnarray}
where $S^{r,m} \equiv \sum_{k=0}^{r}(-1)^{r-k} {k^{m} \over k!(r-k)!}$ 
are the ``Stirling numbers of second kind'' \cite{tabla}.

Comparing Eqs. (\ref{partialalgebra}) and (\ref{partialalgebra1}) or
Eqs. (\ref{multicommutator0}) and (\ref{multicommutator2}) and
looking
for a value of $q$ such that the dynamics of both sets of relevant
operators are isomorphous, we arrive to:

\begin{mathletters}
\label{relations}
\begin{eqnarray}
{\omega_{q} \over \omega_{1}} & = & {n \over [n]_{q}} 
\left(1 + n {\omega_{2} \over \omega_{1}} \right), \\
q(n) & = & { {\omega_{1} \over \omega_{2}}
+ n+2 \over {\omega_{1} \over \omega_{2}} + n }.
\end{eqnarray}
\end{mathletters}

We see from the above relations that $q$ depends on the supra-index $n$
and that $q(n)$ is {\it always greater than one} for all values of $n$ and
the parameters of the system. Note that  $q > 1$ is a {\it necessary condition}
to guarantee that: a) the relevant operators $\hat \Lambda_{q}^{n,m}$ are 
unbounded and b) the multicommutator, associated with the Arik-Coon Hamiltonian, 
can be written as an operator average with respect to the Binomial distribution. 
Therefore, Eqs. (\ref{relations}) assures the same temporal evolutions for, both,
the $q$-deformed and anharmonic oscillator's relevant operators for
a given value of $n$ consistently with a) and b).

The interpretation of the $q$ parameter as a measure of anharmonicity has
been introduced under the conditions for which coherent states of the 
$q$-deformed oscillator and the coherent states of the anharmonic
oscillator
are equivalent \cite{Artoni,Birman}. In our case we look for an
isomorphism which connects the dynamics of both systems {\it without
specifying the initial state}, and we obtain that $q$ is a measure of
anharmonicity which depends on the supra-index $n$. Other interpretation of
the nonlinear aspects of deformed oscillators has been also proposed,
for instance, Mank'o et al. have interpreted the $q$-deformed oscillator
as the usual nonlinear oscillator but with a specific exponential
dependence of
the oscillator frequency on the amplitude of vibration \cite{Manko1}.
Besides they have also given an experimental bound for the value of $q$
\cite{Manko2}.

Summing up, we have studied the dynamical properties of 
$q$-deformed oscillators and their relationship to the anharmonic
oscillators by means of a Lie-algebraic approach. We have found that
an infinite dimensional set of $q$-deformed relevant operators close 
a partial $q$-deformed Lie algebra under commutation with the 
$q$-deformed oscillator Hamiltonian. We have shown that the
dynamical 
properties of the system can be described in terms of the multicommutator 
of the type 
$\lbrack \hat H, \ldots, \lbrack \hat H, \hat O \rbrack \ldots \rbrack$. 
We have also obtained, that the multicommutator can be expressed as a 
power law of $E_{q}(n)\lbrack \hat a_{q}, \hat a_{q}^{\dagger} \rbrack$ and 
that the dynamics of the infinite-dimensional $q$-deformed
Lie-algebra scale.
Finally, we have study differences and similarities between the
$q$-deformed and anharmonic oscillators and we have shown that for 
$|q| < 1$ both models can not be isomorphous, moreover while $\hat \Lambda^{n,m}_{q}$ 
are bounded $\hat \Lambda^{n,m}$ are unbounded, on the other hand 
for $q > 1$ we have proved that the multicommutator for both model can be written as
an operator average with respect to the Binomial distribution and we have
found the value of $q(n)$ which makes the dynamics of both models
isomorphous. Besides, we have found exact expressions for the dynamics
of the relevant operators of both models for a $q$-coherent and coherent states
respectively, and shown the similarity in their functional forms.
We have also concluded that the deformation parameter $q$ can be
interpreted as a measure of anharmonicity which depends on one
of the supra-index (i.e. $n$) of the relevant operators. This last property
leads us to conclude that any magnitude depending on $n$, such as
squeezing, will have a very different temporal evolution
for both systems. 

\section*{Acknowledgements} I would like to thank Prof. J. Katriel
for a critical reading of the manuscript and useful comments.

\end{document}